\documentclass[prd,twocolumn,amssymb,eqsecnum,showpacs,epsfig,altaffilletter]{revtex4}
\usepackage{graphicx}
\usepackage{bm}
\usepackage{dcolumn}
\usepackage{amsmath}

\def \lleq {\lower0.9ex\hbox{ $\buildrel < \over \sim$} ~}
\def \ggeq {\lower0.9ex\hbox{ $\buildrel > \over \sim$} ~}

\def \beq  {\begin{equation}}
\def \eeq  {\end{equation}}
\def \ber  {\begin{eqnarray}}
\def \eer  {\end{eqnarray}}

\begin{document}
\newcommand{\newc}{\newcommand}

\newc{\ga}{\gamma}
\newc{\n}{\label}
\newc{\be}{\begin{equation}}
\newc{\ee}{\end{equation}}
\newc{\ba}{\begin{eqnarray}}
\newc{\ea}{\end{eqnarray}}
\newc{\bea}{\begin{eqnarray*}}
\newc{\eea}{\end{eqnarray*}}
\newc{\D}{\partial}
\newc{\ie}{{\it i.e.} }
\newc{\eg}{{\it e.g.} }
\newc{\etc}{{\it etc.} }
\newc{\etal}{{\it et al.}}
\newcommand{\nn}{\nonumber}
\newc{\ra}{\rightarrow}
\newc{\lra}{\leftrightarrow}
\newc{\lsim}{\buildrel{<}\over{\sim}}
\newc{\gsim}{\buildrel{>}\over{\sim}}
\title{Duality extended Chaplygin cosmologies with a big rip}\date{\today}
\author{Luis P. Chimento}
\email{chimento@df.uba.ar}
\affiliation{
Dpto. de F\'\i sica, Facultad de Ciencias Exactas y Naturales,
Universidad de Buenos Aires, Ciudad Universitaria
Pabell\'on I, 1428 Buenos Aires, Argentina}
\author{Ruth Lazkoz}
\email{ruth.lazkoz
@ehu.es}
\affiliation{Fisika Teorikoa, Zientzia eta Teknologiaren Fakultatea, Euskal Herriko Unibertsitatea, 644 Posta Kutxatila, 48080 Bilbao, Spain}

\begin{abstract}
We consider modifications to the Friedmann equation motivated by recent proposals along these lines pursuing an explanation to the observed late time acceleration. Here we show those modifications can be framed within a theory with self-interacting gravity, where   the term self-interaction refers here to the presence of functions of $\rho$ and $p$ in the right hand side of the Einstein equations. We then discuss the construction of the duals of the cosmologies generated within that framework. After that we investigate the modifications required to generate generalized and modified Chaplygin cosmologies and show that their duals belong to a larger family of cosmologies we call extended Chaplygin cosmologies. Finally, by letting the parameters of those models take values not earlier considered in the literature we show some representatives of that family of cosmologies
display sudden future singularities, which indicates their behavior is  rather different
from generalized or modified Chaplygin gas cosmologies. This reinforces the idea that modifications
of gravity can be responsible for unexpected evolutionary features in the universe.
\end{abstract}
\pacs{98.80.Cq}
\maketitle

\section{Introduction}
All well established cosmological models of the Universe at large scales rest on the basic assumptions that matter is distributed homogeneous and isotropically and that it structure is governed by an effective theory of gravity.

The first assumption implies the large-scale geometry of the Universe is given by a Friedmann-Robertson-Walker metric in which the only non-trivial degree of freedom is  the scale factor $a(t)$, where $t$ is the cosmological time. Hence, observations based on geometry can only tell us about $a$ and/or its derivatives. The second assumption of the two above determines how the geometry evolves. In usual practice the different sources are treated like  fluids with energy density  $\rho_i$ and pressure $p_i$. Essentially, if one knows the total pressure as a function of the total energy density, then the total energy density $\rho$ can be determined  by integrating the corresponding conservation equations. Finally, one can use the gravitational field equations within our effective theory of gravity  to determine the metric. 

General Relativity is the standard effective theory of gravity, but possible corrections to the right hand side (rhs) of the Friedmann equation involving a function of the total energy density $\rho$ have been investigated recently \cite{dgp,dt,car,barsen}.
These modifications have been designed so that they play an important  role in the late asymptotic regimes \footnote{In  brane cosmology, one also ends up having modifications in the rhs of the Friedmann equations involving $\rho$ only when bulk effects are switched off, but those corrections dominate in the early time asymptotic regime}, 
providing an explanation to the currently observed acceleration \cite{sup}. In this fashion one would, only recover Einstein's
gravity, above some energy scales. 



In this paper we show those modified gravity approaches can be framed within a theory with self-interacting gravity, where we use self-interaction to indicate the presence of functions of $\rho$ and $p$ in the right hand side (rhs) of the Einstein equations. 
After that, and within that framework, we show how to apply duality
 transformations \cite{dual} in a similar fashion to previous works regarding conventional scalar field cosmologies or perfect fluid brane cosmologies.  These transformations  have received considerable attention because they lead to peculiar new cosmological models. For instance, the dual of a cosmological model
with an initial singularity or big bang is another model with final big rip singularity
(the duality trades one singularity for the other).

In the spirit of \cite{barsen} we show later on  that Chaplygin cosmologies, which are usually viewed as arising from a fluid with an exotic equation of state, can also be interpreted as stemming from a self-interacting fluid. Specifically, we find a form of the self-interaction which defines a class of cosmologies (first considered in \cite{pedro}) which will call extended Chaplygin cosmologies. This class includes generalized and modified Chaplygin cosmologies (hereafter GC and MC cosmologies) and their duals as 
particular cases.  Interestingly, this class of models displays   a rich casuistics as for the possible evolutions, for instance scenarios with strong and weak sudden future singularities \cite{geo}.  

The plot of the paper is as follows. In section I  we study the equations that govern flat FRW cosmologies
in the setup of self-interacting gravity, and then we show how the duals of those models can be obtained. Then, we investigate which are the kinds of self-interactions leading to GC and MC cosmologies. In section II we show that it is possible to think of a single kind  of self-interaction
which for particular values of the parameters defining it reduces to the previous kinds
of self-interactions used for obtaining GC and MC cosmologies. As we said previously, the family of cosmologies 
obtained from that self-interaction  will be dubbed extended Chaplygin cosmologies.  We then show how to obtain their duals (which happen to belong to the same class of cosmologies). Finally, we concentrate on some of those extended cosmologies which we find particularly appealing, and highlight their properties.

\section{Chaplygin cosmologies within  self-interacting gravity}

We consider now the dynamics of flat FRW cosmologies  governed by the equations of motion
of a gravitational theory with self-interaction in the aforementioned sense:
\ba
3H^2&=&\rho_{s}(\rho),\label{oo}\\
-2\dot H&=&\rho_{s}(\rho)+p_{s}(p)\label{mix},
\ea
where $H=\dot a/a$ is the Hubble factor, $a$ the scale factor, and the dot denotes differentiation with respect to cosmic time.
Here $\rho$ and $p$ are respectively the energy density and the pressure of the fluid that fills the universe. They satisfy the conservation equation 
\be
\dot \rho+3H(\rho+p)=0 \label{cons}.
\ee

Now, from Eqs. (\ref{oo}) and (\ref{mix}) one gets 
\be
\dot \rho_{s}+3H(\rho_{s}+p_{s})=0 \label{cons_self},
\ee
 and combining the latter with Eq. (\ref{cons}) we arrive at
\be
\rho'_{s}=\frac{\rho_{s}+p_{s}}{\rho+p}\label{rel}
\ee
where the prime denotes differentiation with respect to the energy density $\rho$.

We are going to show that  within the  self-interacting gravity
picture GC and MC cosmologies can be constructed in terms of the energy density of the fluid filling the universe.

Let us  assume our self-interacting fluid has a generalized Chaplygin gas equation of 
estate:
\be
p_{s}=-\frac{A}{\rho^{\alpha}_{s}},
\ee
with $\alpha$ and $A$ constants.
Using Eq. (\ref{rel}) for a conventional perfect fluid $p=(\gamma-1)\rho$ with $\gamma>0$ a constant
one arrives at
\be
\rho_{s}=\left(A+\rho^{({1+\alpha})/{\gamma}}\right)^{{1}/{(1+\alpha)}}.
\ee
 This coincides with the result obtained by the modification of gravity proposed in \cite{barsen}, where only the Friedmann equation was modified.

As an alternative, let us now consider   a modified Chaplygin equation of state \cite{luis}:
\be
p_{s}=-\frac{1}{1+\alpha}\left[\alpha\rho_{s}+ \frac{A}{{\rho^{\alpha}_{s}}}\right]\label{modchap}.
\ee
Eq. (\ref{modchap})  mimicks the mixture of a barotropic perfect fluid and a generalized Chaplygin gas. This is actually a purely kinetic k-essence model \cite{arm} which describes the unification of dark matter and dark energy \cite{pedro,sch,inhomo,log1,log2}. 
Using Eq. (\ref{rel})
one gets
\be
\rho_{s}=\left(A+\rho^{1/\gamma}\right)^{{1}/{(1+\alpha)}}.
\ee
MC cosmologies have the merit that, as recently shown in \cite{laznesper}, they   are observationally favored over generalized Chaplygin cosmologies. 
\section{Dual Chaplygin  cosmologies}

A duality transformation links a  cosmology with scale factor $a$ with another one with scale factor $a^{-1}$. In consequence this leads to the transformation $H\to -H$, which
leaves (\ref{oo}) unchanged but reverses the sign of the sum $(\rho_{s}+p_{s})$ in (\ref{mix}). 
Hence $p_{s}$ must  transform according to
\be 
p_{s}\to-(2\rho_{s}+p_{s})\label{trans},
\ee
to leave Eq.(\ref{mix}) unaltered.

Since $\rho_{s}(\rho)$ is invariant under the transformation, so it is its derivative with respect to $\rho$. Combing that with (\ref{rel}) one deduces
that $(\rho+p)\to -(\rho+p)$, so that if the seed fluid satisfies the weak energy condition
the source of the dual cosmology will automatically violate it.

Now, we are going to obtain the  duals  of GC and MC cosmologies taking advantage of the self-interacting gravity concept.
Economy of efforts suggests the  convenience of introducing an equation of
state that includes both GC and MC models as particular cases. This equation of state
is
\be
p_{s}=(\ga_0-1)\rho_{s}-\ga_0\frac{A}{\rho^{\alpha}_{s}}\label{extended},
\ee
with $\ga_0$ and $\alpha$ arbitrary constants. 

Following the same steps as before, and keeping the assumption $p=(\gamma-1)\rho$ combined with the requirement $\alpha\ne-1$ one arrives at
\be
\rho_{s}=\left[A+\rho^{{\ga_0(1+\alpha)}/{\gamma}}\right]^{{1}/{1+\alpha}}.
\label{excha}\ee
The cosmologies derived from the latter were presented in \cite{pedro}, where some aspects of them were discussed. We will dub them as  extended Chaplygin gas cosmologies, and since
they have one additional parameter with respect to GC and MC  cosmologies, they are richer in possible evolutions. The GC and MC cases are obtained by taking in Eq. (\ref{excha})
$\ga_0=1$ and  $\ga_0=1/(1+\alpha)$ respectively. In general, for  positive $\alpha$ and $A$ one has $\rho_{s}\approx A^{\frac{1}{1+\alpha}}$  for  $\rho\approx 0$, and $\rho_{s}\approx\rho^{\ga_0}$
for very large $\rho$ and $\ga_0>0$. In contrast, if $A<0$ consistency requires $\rho$ be bounded from below so that $\rho^{{\ga_0(1+\alpha)}/{\gamma}}\le\vert A\vert$

The application of the 
duality transformation will give us the duals
of all the representatives of the family of extended Chaplygin cosmologies,  GC and MC cosmologies included. The way to obtain the  duals  of these extended Chaplygin cosmologies
is to apply (\ref{trans}) to (\ref{extended}), which is equivalent to the transformations
\ba
&&\ga_0\to-\ga_0.
\ea
If we now denote with the superindex $d$ the quantities obtained under the duality transformation we will have
\be
p_{s}^d=-(\ga_0+1)\rho_{s}+\ga_0\frac{A}{\rho^{\alpha}_{s}}\label{exd},
\ee for the cosmologies which are dual to those obtained from (\ref{extended}).
Note that for the  dual of a GC cosmology one has
\be
p_{s}^d=-2\rho_{s}+\frac{A}{\rho^{\alpha}_{s}}\label{gd},
\ee
so it is not another  GC cosmology, and the same applies to MC cosmologies because for their duals
\be
p_{s}^d=-\frac{1}{1+\alpha}\left[(2+\alpha)\rho_{s}-\frac{A}{\rho^{\alpha}_{s}}\right]\label{md},
\ee
and these correspond to a different subclass of cosmologies.
However the duals of GC and MC cosmologies belong to the larger family of
dual extended Chaplygin cosmologies (\ref{exd}), and since those have an initial singularity due to the vanishing of the scale factor (generically in the GC case and for $\ga_0>0$ in the MC case) their duals will have a sudden future singularity in which the scale factor will blow up in a finite time (see \cite{uniphan} for details). 

Let us now move to analyzing the behavior of some of the new cosmologies we have presented here. Chaplygin cosmologies
were advocated to explain the late time acceleration suggested by supernovae data  \cite{sup}, and we will mainly address in the following the extended Chaplygin cosmologies that can serve that purpose.  Interestingly, the expansion of the Universe seems not to be just accelerated, but super-accelerated ($\dot H>0$ and $H>0$), i.e. the universe can be described as filled with a phantom fluid. Keeping that in mind, we will mainly
care for  the extended Chaplygin cosmologies with late-time super-acceleration, because such asymptotic behavior (which is driven by super-negative pressure) seems to have some motivation in the framework of string theory \cite{string}.
GC and MC cosmologies are typically super-accelerated at late times but as time grows  $\dot H$ becomes smaller and smaller, i.e. they end up being de Sitter cosmologies. In order to highlight the novel aspects of extended Chaplygin cosmologies we will concentrate on those
super-accelerated models which do not have a de Sitter-like late-time limit. Our discussion will remain qualitative, and it will be based on the assumption $p=(\gamma-1)\rho$ which will at the end of the day give us the expression of $\rho_{s}$ in terms of the scale factor.

Extended Chaplygin cosmologies with late-time supper-acceleration and without a de Sitter-like late-time limit
are obtained assuming $\gamma_0<0$ (not considered in the literature so far) together with $1+\alpha>0$,  and two different possibilities arise depending on the sign  of $A$. 

For $A<0$ the scale factor never vanishes and has a lower limiting value $a_0$ where the energy density, as given by (\ref{excha}), vanishes. The initial behavior of the universe corresponds to a contracting phase, and contraction lasts till the scale factor reaches the minimum value $a_0$, then a bounce occurs, and the universe commences to expand and has a final expanding phase. The scale factor evolves between two infinite values. For large values of the scale factor (\ref{excha}) gives $\rho_s\approx a^{3|\ga_0|}$,  which implies $a\approx \vert t-t_i\vert^{-2/3|\ga_0|}$ with $t_i$ representing any of the two time instants where the blow up occurs. The birth and end of this universe are signaled by sudden singularities and the final one has the typical characteristics of a big rip. 
 
In contrast, when $A>0$ the scale factor diverges once in the whole history of the universe, this occurs where the energy density $\rho_s$ diverges. This singularity, which we fix at $t=0$, divides the time axis in two disconnected regions, $t>0$ and $t<0$. Thus, when solving the Einstein equations for this case one gets a scale factor with two disconnected branches, one for each region. In the $t>0$ region, the solution represents a contracting universe, which contracts forever, with an initial infinite scale factor and a vanishing final limit in the far future. In the $t<0$ region, the solution represents an expanding universe which begins to evolve from a vanishing  scale factor and ends in a final big rip at $t=0$. 

Leaving aside super-accelerated cosmologies, there are other extended Chaplygin cosmologies with interesting unusual futures.  The name big rip refers usually to sudden future singularities like those described before in which the scale factor diverges in a finite time. Nevertheless,  recently other sudden future singularities
have been considered in which $H$ blows up in a finite time but $a$ stays finite \cite{sudden}. Extended Chaplygin gases with $\gamma_0<0$ and  $1+\alpha<0$  are examples of that kind of evolution, but they do not display late-time super-acceleration.
Nevertheless, one will only be allowed to say a sudden future singularity is strong only if  $a$ blows up \cite{geo}.

Before closing  it is also worth mentioning  the  alternative equation of state 
\be
p_{s}=-\rho_{s}+\frac{A}{\rho^{\alpha}_{s}}\label{extendedlog},
\ee
with
\be
\rho_{s}=\left[\log\rho^{{A (\alpha+1)}/\gamma}\right]^{{1}/{(1+\alpha)}}.
\label{exchaprho}
\ee
It generates an peculiar class of Chaplygin  cosmologies \cite{log1}, which do not however belong to the 
class defined by the equation of state (\ref{extended}). Comparing the expression (\ref{extendedlog}) with its dual equation of state
\be
p_{s}^d=-\rho_{s}-\frac{A}{\rho^{\alpha}_{s}},
\ee
we see that the duality transformations maps both cosmologies into others within the same class. The evolution of these cosmologies  has been subject of detailed study in \cite{log2}, so we submit the interested reader to this reference.

\section{Conclusions}
In this paper we investigate some possibilities offered by modifications to Einstein's theory of gravity and consequences of them. Our study rests on the customary large-scale description on the universe under the assumption of isotropy and homogeneity.  
Specifically, we address some  modifications to the Friedmann equations which are motivated by recent attempts in this spirit which aim at explaining the observed late time acceleration. 

We propose framing those modified gravity approaches within a theory with self-interacting gravity, where the term self-interaction is used to indicate the presence of functions of $\rho$ and $p$ in the rhs of the Einstein equations. We then anticipate formally how given a cosmology derived from those assumptions one can obtain its dual. This  can have interesting applications of the construction of phantom cosmologies with a big rip taking as seeds more conventional cosmologies with a big bang.

We then apply our proposal to GC and MC cosmologies. First, we discuss the self-interaction needed so that a bare $p=(\gamma-1)\rho$ fluid can act as the source of a GC or MC cosmology. Then we look at the duals of those cosmologies and find they belong to a larger
family of cosmologies which we call extended Chaplygin cosmologies. These have one additional parameter compared to GC or MC cosmologies, and in consequence the collection of 
evolutions they can depict is richer. In particular we concentrate on cases with  not previously considered  values of the parameters and find that these extended Chaplygin cosmologies can have future sudden singularities with the typical big rip (blow up in the scale factor) or rather as those described in \cite{sudden} (regular scale factor). 

\section*{Acknowledgments}
We are grateful to J.M. Aguirregabiria for valuable suggestions.
R.L. is supported by the Spanish Ministry of Science and Education through research grant FIS2004-01626. L.P.C. is partially funded by the University of Buenos Aires  under
project X224, and the Consejo Nacional de Investigaciones Cient\'{\i}ficas y
T\'ecnicas under proyect 02205.


\begin{thebibliography}{99}
\bibitem{dgp}
  G.~R.~Dvali, G.~Gabadadze and M.~Porrati,
  %
  Phys.\ Lett.\ B {\bf 484} (2000) 112;
  G.~R.~Dvali and G.~Gabadadze,
  %
  Phys.\ Rev.\ D {\bf 63} (2001) 065007;
G.~Dvali, G.~Gabadadze and M.~Shifman,
  %
  Phys.\ Rev.\ D {\bf 67} (2003) 044020;
  C.~Deffayet, G.~R.~Dvali and G.~Gabadadze,
  %
  Phys.\ Rev.\ D {\bf 65} (2002) 044023.
\bibitem{dt}G.~Dvali and M.~S.~Turner,
  arXiv:astro-ph/0301510.
\bibitem{car}
K.~Freese and M.~Lewis,
  Phys.\ Lett.\ B {\bf 540} (2002) 1;
  P.~Gondolo and K.~Freese,
  Phys.\ Rev.\ D {\bf 68} (2003) 063509.
\bibitem{barsen}  T.~Barreiro and A.~A.~Sen,
  Phys.\ Rev.\ D {\bf 70} (2004) 124013.
\bibitem{sup}A.~G.~Riess {\it et al.}  [Supernova Search Team Collaboration],
  %
  Astrophys.\ J.\  {\bf 607} (2004) 665.
  \bibitem{string} P.H. Frampton,  arXiv:hep-th/0202063. Phys.\ Lett. B {\bf 555} (2003) 139. 
  \bibitem{dual}
  L.~P.~Chimento and R.~Lazkoz,
  %
  Phys.\ Rev.\ Lett.\  {\bf 91} (2003) 211301;
   M.~P.~Dabrowski, T.~Stachowiak and M.~Szydlowski,
  %
  Phys.\ Rev.\ D {\bf 68} (2003) 103519;
  L.~A.~Boyle, P.~J.~Steinhardt and N.~Turok,
  %
  Phys.\ Rev.\ D {\bf 70} (2004) 023504;
  J.~M.~Aguirregabiria, L.~P.~Chimento and R.~Lazkoz,
  %
  Phys.\ Rev.\ D {\bf 70} (2004) 023509;
Y.~S.~Piao and Y.~Z.~Zhang,
  %
  Phys.\ Rev.\ D {\bf 70} (2004) 043516; 
  Y.~S.~Piao,
  %
  Phys.\ Lett.\ B {\bf 606} (2005) 245;
\,J.~E.~Lidsey,
  %
  Phys.\ Rev.\ D {\bf 70} (2004) 041302;
 G.~Calcagni,
  Phys.\ Rev.\ D {\bf 71} (2005) 023511; 
  arXiv:gr-qc/0410111;
  arXiv:hep-ph/0503044.




\bibitem{bigrip}R.~R.~Caldwell, M.~Kamionkowski and N.~N.~Weinberg,
  Phys.\ Rev.\ Lett.\  {\bf 91} (2003) 071301.
  \bibitem{pedro}
  P.~F.~Gonzalez-Diaz,
  Phys.\ Rev.\ D {\bf 68} (2003) 021303.
\bibitem{geo}
L.~Fernandez-Jambrina and R.~Lazkoz,
  Phys.\ Rev.\ D {\bf 70} (2004) R121503.
\bibitem{arm}
  C.~Armendariz-Picon, T.~Damour and V.~Mukhanov,
  Phys.\ Lett.\ B {\bf 458} (1999) 209.
\bibitem{sch}R.~J.~Scherrer,
  Phys.\ Rev.\ Lett.\  {\bf 93} (2004) 011301 .
\bibitem{inhomo}L.~P.~Chimento and R.~Lazkoz,
  arXiv:astro-ph/0411068.
\bibitem{log1}D.~J.~Liu and X.~Z.~Li,
  arXiv:astro-ph/0501115.
  \bibitem{log2}
H.~Stefancic,
 Phys.\ Rev.\ D {\bf 71} (2005) 084024.
\bibitem{laznesper}R.~Lazkoz, S.~Nesseris and L.~Perivolaropoulos,
  arXiv:astro-ph/0503230.
  \bibitem{uniphan}
  L.~P.~Chimento and R.~Lazkoz,
  arXiv:astro-ph/0405518.
\bibitem{sudden}
  J.~D.~Barrow,
  Class.\ Quant.\ Grav.\  {\bf 21} (2004) L79,
  Class.\ Quant.\ Grav.\  {\bf 21} (2004) 5619;
  L.~P.~Chimento and R.~Lazkoz,
  Mod.\ Phys.\ Lett.\ A {\bf 19} (2004) 2479;
S.~Nojiri and S.~D.~Odintsov, 
  Phys.\ Lett.\ B {\bf 595} (2004) 1;  Phys.\ Rev.\ D {\bf 70} (2004) 103522 ;
 K.~Lake,
  Class.\ Quant.\ Grav.\  {\bf 21} (2004) L129;
  Class.\ Quant.\ Grav.\  {\bf 22} (2005) L35;
  M.~P.~Dabrowski,
  arXiv:gr-qc/0410033.
 J.~D.~Barrow and C.~G.~Tsagas,
 Class.\ Quant.\ Grav.\ {\bf 22} (2005) 1563;
\bibitem{luis} 
L. P. Chimento, Phys. Rev. D {\bf 69} (2004) 123517.
  \end{thebibliography}
\end{document}